# Local determination of the constitutive law of a dense suspension of non-colloidal particles through MRI

Guillaume Ovarlez[*], François Bertrand, Stéphane Rodts
Laboratoire des Matériaux et Structures du Génie Civil (UMR 113 LCPC-ENPC-CNRS)
Institut Navier - 2, allée Kepler, 77420 Champs-sur-Marne, France

August 9, 2005

**Synopsis**

We investigate the flowing behavior of dense suspensions of non-colloidal particles, by coupling macroscopic rheometric experiments and local velocity and concentration measurements through MRI techniques. We find that the flow is localized at low velocities, and that the material is inhomogeneous; the local laws inferred from macroscopic rheometric observations must then be reinterpreted in the light of these local observations. We show that the short time response to a velocity step allows to characterize dense suspensions locally: they have a purely viscous behavior, without any observable influence of friction. In the jammed zone, there may be a contact network, whereas in the sheared zone there are only hydrodynamic interactions: localization consists in a change in configuration at the grain scale. From the concentration and velocity profiles, we have provided for the first time local measurements of the concentration dependence of viscosity; we find a Krieger-Dougherty law $\eta(\phi) = \eta_0(1 - \phi/0.605)^{-2}$. Shear induced migration is almost instantaneous and seems inconsistent with most observations: it would imply that the diffusion coefficients strongly depend on the concentration. We finally propose a simple constitutive law for dense suspensions, based on a purely viscous behavior, that accounts for all the macroscopic and local observations.

## I  Introduction

The flows of concentrated suspensions are involved in many industrial processes (coal slurries transport, sewage sludge application, foodstuff transport...), natural phenomena (mud flows, lava flows, landslides...) and in daily use (cosmetic pastes spreading, painting...). It is of high importance to predict and optimize the flow behavior of these materials: drilling muds, e.g., have to carry debris back out of drill holes; in fresh concrete, a challenge is to pack as many solid particles as possible together in a given volume, so that the solid concrete will be resistant, while having a sufficiently fluid material in order to facilitate its transport and its placing. In the case of natural flows, one wants to predict the extent of the flows. However, the flows of concentrated suspensions reveal many complex features which are far for being understood (for a recent review, see Stickel and Powell (2005)). This complexity originates from the great variety of interactions between the particles (colloidal, hydrodynamic, frictional, collisional) and of physical properties of the particles (packing fraction, deformability, sensitivity to thermal agitation, shape, buoyancy...) involved in the material behavior. The link between the relative importance of the different kinds of interactions and the different flow regimes have been clarified by Coussot and Ancey (1999).

---

[*]corresponding author: guillaume.ovarlez@lcpc.fr

In this paper, we will focus on the behavior of suspensions of non-colloidal rigid particles suspended in a Newtonian fluid. Therefore, the macroscopic behaviors originating from Brownian motion, colloidal interactions, and deformability of the particles will not be considered.

The viscous flows of non-colloidal suspensions have been extensively studied. Of great importance is the knowledge of the concentration dependence of viscosity, as one may want to transport suspensions as concentrated as possible. For a dilute suspension, it was shown by Einstein (1956) that the viscosity $\eta(\phi)$ may be written as: $\eta(\phi) = \eta_0(1 + 2.5\phi)$, where $\phi$ is the concentration and $\eta_0$ is the viscosity of the interstitial fluid. This relationship was extended to second order by Batchelor and Green (1972) who found: $\eta(\phi) = \eta_0(1 + 2.5\phi + 7.6\phi^2)$. These two theoretical expressions are well verified experimentally for $\phi < 10\%$. For more concentrated suspensions, there is actually no theoretical nor experimental consensus on the concentration dependence of viscosity. The only universal feature which is observed is that the viscosity diverges at the approach of a maximum packing fraction $\phi_m$ which is highly dependent on the particle characteristics. Many phenomenological models have been proposed, the most famous being those by Eilers (1941), Mooney (1951), and Krieger and Dougherty (1959). Basically, all these models intend to recover the Einstein limit at low concentration, and to account for the divergence at $\phi = \phi_m$. The now classical Krieger-Dougherty law reads

$$\eta(\phi) = \eta_0(1 - \phi/\phi_m)^{-n} \qquad (1)$$

In order to recover the Einstein limit at low concentration, one should set $n = 2.5\phi_m$. However, it appears that setting $n$ as a free parameter allows to fit many experimental data; nevertheless, the values of the maximum packing fraction $\phi_m$ and exponent $n$ are still subject to intense debate, since experimental results may differ, particularly at the approach of the maximum packing fraction. These parameters actually depend on the particle shape (see e.g. Barnes *et al.* (1989)), and the polydispersity [Chong *et al.* (1971)], but other factors, mainly measurement problems such as shear induced migration and shear induced resuspension which will be detailed thereafter, may cause the high scattering of data for similar systems. Much theoretical work attempted to understand the concentration dependence of viscosity for very dense suspensions. Frankel and Acrivos (1967) showed that the energy dissipation in suspensions originates mainly from the fluid sheared in the small zones between close particles: one has to consider the lubricated flow between particles; this causes the divergence of viscosity as the particles come closer to each other (i.e. as $\phi$ tends to $\phi_m$). Marrucci and Denn (1985) then showed that the functional form of this divergence depends strongly on the particle configuration; dissipation is much more important when large scales structures, where the gaps between particles are much thinner than the average gap, develop, than when the particles are on a regular lattice. For a detailed discussion on measurements of suspension viscosity, on theoretical models and numerical simulations, see e.g. Barnes *et al.* (1989) and the review by Stickel and Powell (2005).

However, defining a viscosity in these systems may be problematic as dense suspensions exhibit non-newtonian behaviors [Stickel and Powell (2005)]: basically, one may observe the existence of a shear thinning behavior at low shear rates, i.e. the apparent viscosity $\eta(\dot{\gamma})$ is a decreasing function of the shear rate $\dot{\gamma}$, and shear thickening behavior at high shear rates, i.e. $\eta(\dot{\gamma})$ is an increasing function of $\dot{\gamma}$; as a consequence, the material is macroscopically viscous in a limited range of shear rates. For the sake of consistency, many authors choose to study the concentration dependence of viscosity in the zero shear rate limit (provided there is no yield stress so that there is a viscosity plateau at very low shear rates) or in the high shear rate limit (before the onset of shear thickening).

At high shear rates, above a critical shear rate which depends on the particle and fluid parameters, the viscosity of concentrated suspensions is found to increase strongly with the shear rate: this is a reversible shear thickening behavior (for a review, see Barnes (1989)). This phenomenon may have at least 2 origins: a first one, suggested by Bagnold (1954), is that there is a transition from a viscous flow to a collisional flow: viscous dissipation in the interstitial fluid may be replaced by dissipation in grain-grain collisions, which leads to a shear stress proportional to the square of the shear rate (although laminar-turbulent transition may occur in the viscous suspension before this regime can be reached [Hunt *et al.* (2002)]). A second one, evidenced experimentally by Hoffman (1972) (although in colloidal suspensions), is that there is a reversible transition from a two-dimensional layered arrangement of particles to a random three-dimensional form: in this case, the viscosity increases as a consequence of the high dependence of viscosity on configuration evidenced e.g. by Marrucci and Denn (1985). As pointed out by Barnes (1989), these two explanations are not contradictory: the collisional regime may exist for low interstitial fluid viscosity and high particle inertia.

At low shear rates, the viscosity of dense suspensions is found to decrease with the shear rate. Actually, very dense suspensions exhibit a static yield stress: it was evidenced by Husband *et al.* (1993) in creep tests for concentrations $\phi > 0.47$, and by Huang *et al.* (2005) in viscosity bifurcation experiments. Huang *et al.* (2005) also showed that this yield stress is associated with a viscosity bifurcation: above the critical shear stress $\tau_c$, the shear rate is higher than a critical shear rate $\dot{\gamma}_c$; below $\dot{\gamma}_c$, no steady flow exists. The main characteristics of the transition between the shear thinning behavior and the viscous behavior have been pointed out by various authors [Bagnold (1954); Prasad and Kytömaa (1995); Ancey and Coussot (1999); Ancey (2001); Huang *et al.* (2005)]: it has been found (i) that at low shear rates, the shear stress is almost constant as in sheared dry granular materials, (ii) that, as in dry granular materials, the shear stress is proportional to the normal stress (measured on the internal cylinder in a Couette experiment [Bagnold (1954)], on the top plate in a plane shearing experiment [Prasad and Kytömaa (1995)], or postulated to be proportional to the height of the sheared material [Ancey and Coussot (1999)]), (iii) that at high shear rates, the shear stress is proportional to the shear rate and to the viscosity of the interstitial fluid. An interpretation of the microscopic origin of the two flow regimes, based on the results from macroscopic rheometric experiments has been proposed by these authors: the generally accepted picture for the flows of these dense suspensions is that, depending on the shear rate and on the viscosity of the interstitial fluid, the material may either behave similarly to dry granular systems with dominant frictional contacts between the grains (low shear rate, low viscosity), or to viscously dominated systems, when the contacts are lubricated (high shear rate, high viscosity); at very high shear rates, the flow may then become collisional [Bagnold (1954); Savage and McKeown (1983)], although laminar-turbulent transition may occur in the viscous suspension before this regime can be reached [Hunt *et al.* (2002)]. In this picture, the origin of the apparent yield stress is friction: one may thus characterize the onset of flow by a static coefficient of friction rather than by a yield stress (i.e. by a Coulomb criterion rather than by a Von Mises criterion). The flow behavior of dry granular matter has been the subject of considerable work recently, and a rather good though not simple understanding of the constitutive law in these systems has been reached (see GDR Midi (2004) for a recent review): it would then be possible to compare accurately the low shear rate flow regime of dense suspensions to dry granular material flows. Actually, in a recent study of dense suspension flows, Huang *et al.* (2005) have evidenced shear localization at low shear rates, and showed that the velocity profiles in the flows of dense suspensions are very similar to those observed in dry granular material flows.

The normal stresses that develop under shear were extensively studied by Zarraga *et al.* (2000). It appears that, unlike what happens in polymers, the normal stresses are proportional to the shear rate, for concentrations as low as 30%; this is in agreement with the observations of [Bagnold (1954); Prasad and Kytömaa (1995); Ancey and Coussot (1999); Huang *et al.* (2005)]. The $\dot{\gamma}$ dependence of normal stresses originates from the hydrodynamic interactions, whereas the emergence of normal stresses difference may be related to an anisotropy in the microstructure that is generated by hard spheres repulsion [Brady and Morris (1997)]. An important implication is that a constant ratio between normal and shear stresses is not specific to granular flows: such a proportionality arises when shear stress and a normal stress are both proportional to the shear rate. Therefore, one may have to be cautious before considering the observation of a constant ratio between normal and shear stresses as a proof of the existence of frictional dissipation.

Up to now, the main features of the different flow regimes have thus been identified, and their physical origin is now partly understood, but the construction of a constitutive law of dense suspensions valid over a wide range of shear rates (i.e. accounting for all the observed phenomena) is far from being achieved.

Nevertheless, so far, all the results on the behavior of dense suspensions and its microscopic origin were inferred from macroscopic rheometric experiments. Performing good measurements in dense suspensions is however very difficult as many perturbation effects may develop during the tests: the most severe, which may have caused the high scattering of data observed over years, are wall slip effects [Barnes (1995); Jana et al. (1995)], shear localization [Coussot (2005)] and particle migration.

The migration phenomenon, which leads to concentration inhomogeneities, and is at the origin of size segregation in polydisperse systems, was observed in many situations: Couette flows [Leighton and Acrivos (1987b); Graham et al. (1991); Abbott et al. (1991); Phillips et al. (1992); Chow et al. (1994); Corbett et al. (1995); Tetlow et al. (1998); Shapley et al. (2004)], parallel-plate flows [Chow et al. (1994); Barentin et al (2004)], pipe flows [Sinton and Chow (1991); Altobelli et al. (1991); Hampton et al. (1997); Lyon and Leal (1998); Butler et al. (1999); Han et al. (1999)], and extrusion experiments [Altobelli et al. (1997); Götz et al. (2002)]. In the Couette flows, the consequence of migration is an excess of particles near the outer cylinder. Migration seems to be related to the shear-induced diffusion phenomenon [Acrivos (1995)]. Shear induced diffusion was investigated experimentally [Leighton and Acrivos (1987a); Breedveld et al. (1998, 2002)], theoretically [Morris and Brady (1996); Brady and Morris (1997)], and in simulations [Drazer et al. (2002)]. Basically, the diffusion coefficient $D$ may be written as [Leighton and Acrivos (1987a); Acrivos (1995)]:

$$D = \bar{D}(\phi)\dot{\gamma}a^2 \qquad (2)$$

where $\phi$ is the volume fraction, $\dot{\gamma}$ the shear rate, $a$ the particle size, and $\bar{D}(\phi)$ is the dimensionless diffusion coefficient, which may be a tensor. In the diffusive model of Leighton and Acrivos (1987b) and Phillips et al. (1992), the gradients in shear rate that exist in all but the cone and plate geometry then generate a particle flux towards the low shear zones (i.e. the outer cylinder in the case of the Couette geometry), which is counterbalanced by a particle flux due to viscosity gradients. A steady state, which results from competition between both fluxes, may then be reached, and it is characterized by an excess of particles in the low shear zones of the flow geometry. There are many other models, such as the Morris and Boulay (1999) model in which particle fluxes counterbalance the gradients in normal stresses; most of them were recently compared to experimental data by Shapley et al. (2004).

For slight differences in density between the particles and the suspending fluid, when one may think that sedimentation effects are negligible, one also observes a small dependence of the macroscopic viscosity on the shear rate: this may be attributed to a vertical inhomogeneity of the suspension, which results from competition between shear induced resuspension [Gadala-Maria and Acrivos (1980); Leighton and Acrivos (1986); Acrivos et al. (1993, 1994); Acrivos (1995); Tripathi and Acrivos (1999)], whose origin is again shear induced diffusion, and sedimentation; it results in a inhomogeneity dependent on the shear rate, that creates an apparent shear-thinning behavior.

However most experiments [Graham et al. (1991); Phillips et al. (1992); Corbett et al. (1995)] observe that the migration phenomenon in suspensions of volume fraction up to 55% is rather slow, in accordance with its diffusive origin: it lasts for a few thousands revolutions. Therefore, it is believed that if one performs viscosity measurements on a dense suspension of mean volume fraction $\phi$ at the beginning of shear, then these measurements are performed on a homogeneous suspension and thus provide the value of the viscosity at a concentration $\phi$.

In this paper, we focus on the behavior of a model dense suspension of monodisperse spherical particles at low and intermediate shear rates, i.e. on the shear-thinning and viscous regimes; we do not deal with the shear thickening regime that emerges at high shear rates. Our aim is to determine the shear part of the dense suspensions constitutive law (we do not deal with normal stresses which were extensively studied by Zarraga *et al.* (2000)), and to relate it to the physical properties of the material. We perform rheometric experiments; inferring a constitutive law from macroscopic data may however lead to serious misinterpretation if e.g. the shear flow and the material are inhomogeneous (the shear rate and concentration would then be wrongly estimated). Therefore, in order to interpret carefully the macroscopic data, we measure the local velocity and concentration through MRI techniques: we are then be able to base our determination of the constitutive law and its dependence on the volume fraction on the true local shear rate and the true local concentration. In Sec. II, we present the experimental display. We present the experimental results (rheometric data, velocity and concentration profiles) obtained during the steady flows in Sec. III A, and the transient response to a velocity step in Sec. III B. We analyze the results in Sec. IV: we show how it is possible to determine locally the constitutive law of the material and its local dependence on concentration; the concentration profiles are then compared with previous experimental results and with the predictions of theoretical models. Finally, we propose a simple constitutive law in Sec. V and show how it may account for the macroscopic and local behaviors observed in our experiments.

## II   Materials and methods

We study the rheological behavior of a dense suspension composed of non-colloidal monodisperse spherical particles immersed in a Newtonian fluid at a volume fraction $\phi$ between 55 and 60%; most results presented here are obtained at $\phi = 58\%$, but the observations and conclusions are general as all the suspensions studied exhibited the same features. The particles are monodisperse spherical polystyrene beads (diameter 0.29mm $\pm$ 0.03mm, density 1.05g cm$^{-3}$). The suspensions are prepared with Rhodorsil silicone oil of viscosity 20mPa s, and of density 0.95g cm$^{-3}$. This system allows to change easily the viscosity of the interstitial fluid (the influence of this parameter was studied by Huang *et al.* (2005)) and to match closely the densities of the beads and the fluid in order to avoid sedimentation effects, and also allows to refer to the results obtained by Huang *et al.* (2005) on the same system. We checked by MRI measurements that sedimentation is indeed negligible during the experiments so that there is no dependence of concentration on the height in the gap, except for edge effects.

The rheometric experiments are performed within a Couette geometry (inner cylinder radius $R_i = 4.15$cm, outer cylinder radius $R_e = 6$cm, height $H = 11$cm) on a commercial rheometer (Bohlin C-VOR 200) that imposes either the torque or the rotational velocity (with a torque feedback). In order to avoid slip at the walls, sandpaper of roughness equivalent to that of the particles is glued on the walls; we could check on the velocity profiles that there is no observable slip.

In the rheometric experiments presented here, we control the rotational velocity of the inner cylinder, for values ranging from 0.01 to 100rpm, and we record the torque exerted by the material on the inner cylinder.

Throughout this report, we choose to present the rheometric data as torque measurement vs. rotational speed, because shear rates and stresses would denote an incorrect interpretation of data if the flow is inhomogeneous and if the shear stress on the cylinder depends on height (as in [Ancey and Coussot (1999)]).

Proton MRI [Callaghan (1991)] was chosen as a non-intrusive technique in order to get instant measurements of the local velocity and of the local bead concentration inside the sample. Experiments are performed on a Bruker 24/80 DBX spectrometer equipped with a 0.5T vertical superconductive magnet with 40cm bore diameter and operating at 21MHz (proton frequency).

Its birdcage radio frequency coil is 20cm in size (inner diameter), and its gradients coils are able to deliver field gradients up to 5G/cm with a 500$\mu$s raising time.

We perform our experiments with a home made NMR-compliant rheometer, specially designed to work inside the magnet, with the same Couette geometry as the rheometric experiments. This device was already involved in a number of previous rheo-nmr studies [Raynaud *et al.* (2002); Coussot *et al.* (2002b); Huang *et al.* (2005); Rodts *et al.* (2005)], and is fully depicted in [Raynaud *et al.* (2002)].

The volume imaged is a (virtual) rectangular portion of 40mm in the axial direction with a width (in the tangential direction) of 10mm and a length of 70mm (in the radial direction, starting from the central axis). This volume is situated at the magnet center (so as to damp the effects of field heterogeneities) and sufficiently far from the bottom and the free surface of the rheometer so that flow perturbations due to edge effects remain negligible. We checked that the velocity and concentration profiles are homogeneous along the vertical axis so that we do not perform averages over different profiles in this slice.

Details on the sequence used to obtain velocity profiles can be found in [Raynaud *et al.* (2002); Rodts *et al.* (2004)]. Note that measurements are performed on the oil phase. In all experiments, we measure quasi-instantaneous velocity profiles (a single measurement lasts for 2.3 s). The inner cylinder is driven at velocity $\Omega_i$ ranging from 0.01 to 100rpm. For technical reasons, we can indeed choose $\Omega_i$ either between 0.01 and 9rpm, or between 1 and 100rpm. Therefore preshear at high rotational speed cannot be applied in all experiments.

The NMR sequence used in this work for the measure of the local bead concentration is inspired from sequences aiming at measuring velocity profiles along one diameter in Couette geometry [Hanlon *et al.* (1998); Raynaud *et al.* (2002)]. It consists in a spin-echo sequence, where space selective pulses are used in order to virtually cut in y and z direction a beam along one diameter of the rheometric cell. According to basic MRI spin-warp technique, NMR data are recorded during a so-called read-out gradient, and then Fourier transformed, so as to get 1D information about hydrogen density along x direction. Slight Gaussian filtering is applied in order to remove spurious oscillation of NMR profile close to the edge of the sample. The sequence echo time is 6ms. This is orders of magnitude shorter than the typical time for the inner cylinder to turn around the vertical axis of the cell, and any measurement artefact due to the flowing of the sample during the sequence could thus be neglected; we actually checked on the oil alone that the profiles obtained on the flowing material are independent of the velocity and are the same as the profile obtained on the resting material, as long as the rotational velocity is less than 25rpm. During measurements, due to both physical properties of the sample components and of the Couette device, only NMR signal originating from those hydrogen nuclei belonging to the liquid phase of the sample is recorded. Raw NMR profiles are thus actually representative of the liquid content profile inside the selected beam along x direction. Because of space inhomogeneity of the NMR coil response throughout the sample, a calibration step is necessary in order to make such measurements quantitative (see for instance Graham *et al.* (1991); Corbett *et al.* (1995) for the same kind of application). For this purpose, reference data are taken with the rheometric cell only filled with pure oil. We then compare the NMR signal intensity $S_{\text{oil}}(R)$ at radius $R$ measured in the pure oil, with the NMR signal intensity $S_{\text{suspension}}(R)$ at radius $R$ measured in the suspension in the same conditions. The discrepancy between both signal intensities is attributed to the presence of beads, which give no signal. During experiments, the bead volume fraction $\phi(R)$ inside the suspension is then eventually estimated as:

$$\phi(R) = \frac{S_{\text{oil}}(R) - S_{\text{suspension}}(R)}{S_{\text{oil}}(R)} \quad (3)$$

T1 and T2 relaxation times in the pure fluid were both measured at about 500 ms, and remained unchanged with the presence of beads, so that no further relaxation correction of the data is necessary. A rather low absolute uncertainty of $\pm 0.2\%$ on the concentration measurements values could be estimated from repeated experiments: this ensures that the quantitative evolution of the concentration (with the radius or time) in an experiment is measured accurately. However, there still is a systematic uncertainty in the concentration measurements, mainly due to the fact that the reference measurement (on the oil phase) and the profiles on the various suspensions are performed on different days, and we can never be sure that the spectrometer tunings are exactly the same. We found NMR signal intensity variations of about 1% from one day to another; therefore, there is a systematic uncertainty on the concentration measurements of about ±0.5% (it would then affect in the same way a whole profile as all the data of a given profile are obtained in the same conditions). When a same suspension is studied on two different days, if necessary, we correct the intensity so as to ensure consistency between the signals measured on the suspension at rest on both days. We plan to study in more detail all the sources of systematic uncertainties and we will try to correct them when possible.

Note finally that NMR did not allow at this stage for measurements near the edges: we could not access the volume faction in a 2-3mm zone near the inner and outer cylinders.

## III  Experimental results

### A  Steady state

In this section, we apply a constant rotational velocity, and we record the torque, the velocity profile, and the concentration profile when a steady state is reached. We then study the evolution of the stationary torque and profiles with the rotational velocity, for dense suspensions of volume fraction ranging from 55 to 60%. We study velocities ranging from 0.01rpm to 100rpm. The transient states will be studied in Sec. III B.

#### 1  Torque measurements

In Fig. 1, we plot the torque vs. the rotational velocity for the steady flows of a 58% suspension.

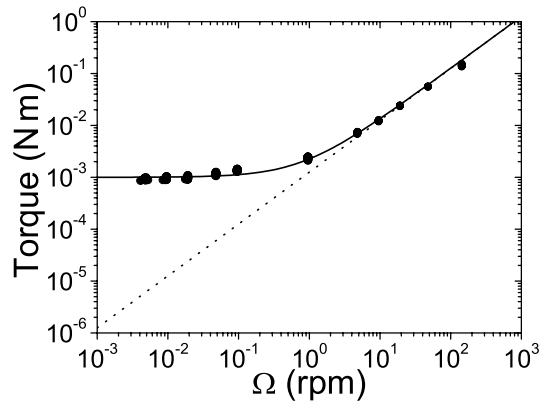

Figure 1: Torque vs. the rotational velocity in the steady state of a 58% suspension. The line is a fit to a Bingham model: $T = T_0 + \alpha\Omega$ with $T_0 = 0.001\,\mathrm{N\,m}$ and $\alpha = 0.012\,\mathrm{N\,m/s}$; the dotted line is a fit to a viscous model $T = \alpha\Omega$ with $\alpha = 0.012\,\mathrm{N\,m/s}$.

As usually observed in dense suspensions (see introduction), we observe a shear-thinning regime followed by a viscous regime, i.e. we find a shear torque plateau at low velocities and a linear increase of torque with the velocity above a critical velocity $\Omega_c$ (Fig. 1). The data can be well fitted to a Bingham law $T = T_0 + \alpha\Omega$ with $T_0 = 0.001\,\mathrm{N\,m}$ and $\alpha = 0.012\,\mathrm{N\,m/s}$: these materials seem to exhibit a yield stress and to behave macroscopically like yield stress fluids.

The rheograms obtained for the other volume fractions studied (55 to 60%) present the same features. In the following, the regime for which the torque increases linearly with the rotational velocity will be called the 'macro-viscous' regime (as in [Bagnold (1954)]), in order to make a clear distinction between the macroscopic observation performed in a rheometric experiment and its interpretation as the constitutive law of the material.

In a simple rheometric experiment, without any other information about what happens inside the gap, one may measure a macroscopic viscosity $\eta(\phi)$ under the assumption that the flow is Newtonian and homogeneous: it allows to infer the shear rate from the rotational velocity, and to define $\eta(\phi)$ for the material of mean volume fraction $\phi$. In the macro-viscous regime, we infer the concentration dependence of the macroscopic viscosity $\eta(\phi)$ of the suspension on concentration from the rheograms obtained for various volume fractions $\phi$ ranging from 55 to 60%. In a torque $T$ vs. rotational velocity $\Omega$ plot, we simply fit the data for high velocities to an asymptotic law $T = \alpha(\phi)\Omega$, and obtain the dimensionless macroscopic viscosity $\eta(\phi)/\eta_0 = \alpha(\phi)/\alpha(0)$, where $\eta_0$ is the suspending fluid viscosity.

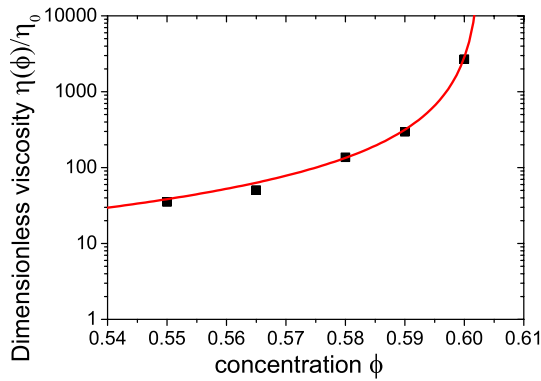

Figure 2: Dimensionless macroscopic viscosity inferred from the macro-viscous regimes of dense suspensions vs. the mean volume fraction of the suspensions. The line is a fit to a Krieger-Dougherty law: $\eta(\phi)/\eta_0 = (1 - \phi/\phi_m)^{-n}$ with $\phi_m = 0.603$ and $n = 1.5$. These data were obtained for suspensions of volume fraction ranging from 55 to 60%.

We observe in Fig. 2 that the macroscopic viscosity increases strongly as the volume fraction increases: there is a factor one hundred in macroscopic viscosity between the 55% suspension and the 60% suspension. In this limited concentration range, the evolution of the dimensionless macroscopic viscosity with the concentration can be well fitted to a Krieger-Dougherty law $\eta(\phi)/\eta_0 = (1 - \phi/\phi_m)^{-n}$ of maximum packing fraction $\phi_m = 0.603$ and divergence exponent $n = 1.5$.

## 2 Velocity profiles measurements

In Fig. 3, we plot the dimensionless velocity profiles for the steady flows of a 58% suspension, for various rotational velocities ranging from 0.07 to 100rpm, i.e. the experimental conditions are the same as for the data of Fig. 1.

MRI measurements show that the velocity profiles are roughly exponential, as in dry granular materials, and that they occupy only a small fraction of the gap at low rotation velocities (Fig. 3): we observe shear localization. In the shear-thinning regime ($\Omega < \Omega_c$), upon increasing the rotation velocity, we find that the higher the rotation velocity, the larger the fraction of the paste that is sheared. Beyond $\Omega_c$, in the case of the 58% suspension, the whole sample is sheared, and all the reduced velocity $V(R)/V(R_i)$ profiles plotted vs. the radius fall along the same curve

(Fig. 3):

$$V(R) = \Omega R_i f(R) \tag{4}$$

However, the velocity profiles in the macro-viscous regime are very different from those expected for a Newtonian fluid; for a Newtonian fluid, we would expect:

$$\frac{V(R)}{V(R_i)} = \frac{R_i}{R} \frac{R_e^2 - R^2}{R_e^2 - R_i^2} \tag{5}$$

i.e a quasi linear velocity profile (Fig. 3). As evidenced by Huang *et al.* (2005), the critical velocity below which the shear flow is localized is the same as the one below which we observe a torque plateau in Fig. 1 (i.e. $\Omega_c \approx 5$rpm). Note that these first results (the velocity profiles in the steady state of a 58% suspension) were already presented in [Huang *et al.* (2005)], and that localization in dense suspensions was also evidenced by Barentin et al (2004).

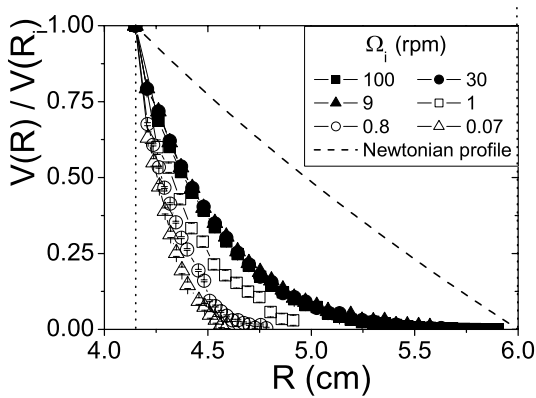

Figure 3: Dimensionless velocity profiles (from Huang *et al.* (2005)) in the steady state of a 58% suspension, at various rotational velocities ranging from 0.07 to 100rpm; the dashed line is the theoretical dimensionless velocity profile for a Newtonian fluid.

It was also shown by Huang *et al.* (2005) that in both regimes, all the dimensionless velocity data $V(R,\Omega)/\Omega R_i$ can be collapsed onto the same universal curve provided they are plotted as a function of the rescaled coordinate $(R - R_i)/d_c(\Omega)$. The length $d_c(\Omega)$, which is an increasing function of $\Omega$ for $\Omega < \Omega_c$, simply gives the thickness of the sheared layer, and is thus constant and equal to the gap size if $\Omega > \Omega_c$. In Sec. III B, we will use this scaling to evaluate the extent of the sheared layer (see e.g. Fig. 9).

As a summary, we observe at low velocities that the velocity profiles are similar to those obtained in dry granular materials, with a roughly constant torque in both cases, as would be expected for a frictional flow, but the thickness of the sheared layer is constant (5-10 grains) in the case of dry granular materials flows [GDR Midi (2004)] in the same range of shear rates, which is a major difference; moreover, when all the material is sheared, at high velocities, the torque is now proportional to the rotational velocity, as would be expected for a viscous flow, but the velocity profile is non-Newtonian. We will explain these paradoxical features in Sec. IV.

In Fig. 4, we plot the dimensionless velocity profiles in a 59% suspension, for velocities ranging from 10 to 30rpm. These rotational velocities belong to the macro-viscous regime of the 59% suspension: the torque is proportional to the rotational velocity (as in Fig. 1) and we observe in Fig. 4 that the dimensionless velocity profiles $V(R)/\Omega_i R_i$ fall along the same curve for all values of $\Omega_i > \Omega_c$. In the case of a suspension of mean volume fraction 58%, we have seen that in the macro-viscous regime all the gap is sheared. In the case of the macro-viscous regime of the 59% suspension, we now observe that the gap cannot be fully sheared (Fig. 4): there still is a 3-4mm region near the outer cylinder where the material is not sheared. When the rotational velocity increases, the thickness of the sheared layer does not increase anymore for $\Omega_i > \Omega_c$, and it remains smaller than the gap size.

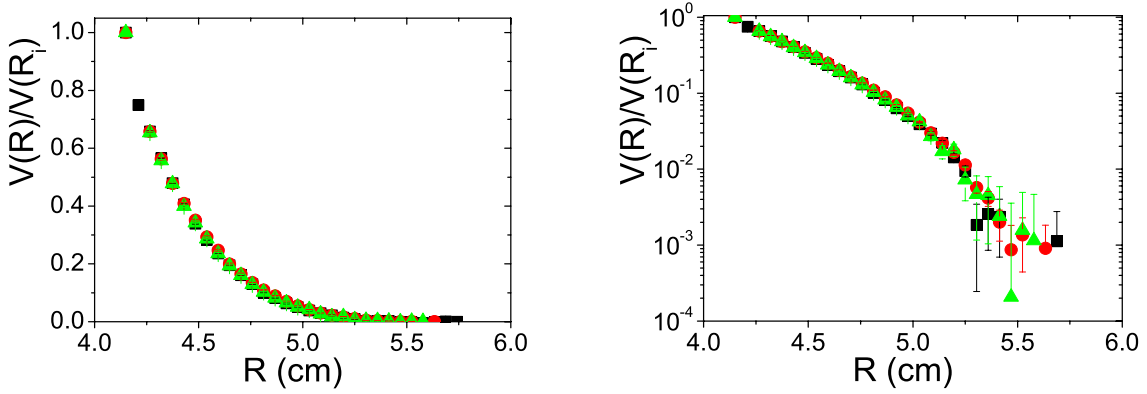

Figure 4: Dimensionless velocity profiles at 30 (squares), 20 (circles), and 10rpm (triangles) for a 59% suspension (lin and log scales).

In Fig. 5, we plot the dimensionless velocity profiles for 58, 59 and 60% suspensions sheared at 20rpm. This 20 rpm velocity belongs to the macro-viscous regime of these 3 suspensions. We now see in Fig. 5 that for the 60% suspension, the material is sheared up to around 5.1cm.

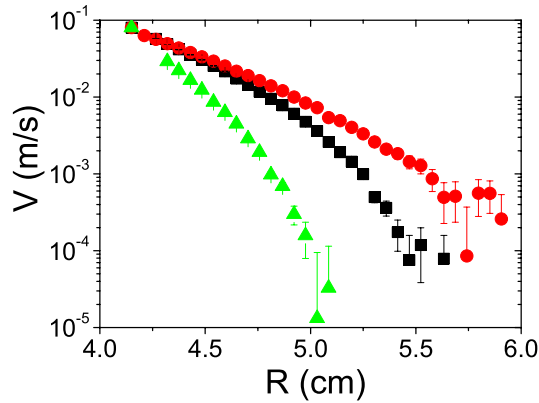

Figure 5: Velocity profiles obtained at 20rpm for 58% (circles), 59% (squares) and 60% (triangles) suspensions.

We have thus observed that the thickness of the sheared layer in the macro-viscous regime (i.e. for $\Omega_i > \Omega_c$) of a dense suspension, in our Couette geometry, does not depend on the velocity and cannot exceed a maximum value $d_m$. This value $d_m$ decreases when the volume fraction increases. A value of $d_m = 1.5$cm was found for the 59% suspension, whereas $d_m = 1$cm for the 60% suspension. In the case of the 58% suspension we can only say that $d_m \geq 1.85$cm. Note however, that this localization, which occurs when increasing the volume fraction in the macro-viscous regime, should not be mistaken for the localization observed in Fig. 3, which occurs when decreasing the velocity below $\Omega_c$ for a given volume fraction. We will show in Sec. IV that these two phenomena may have two different origins.

## 3 Concentration profiles measurements

In Fig. 6, we plot the concentration profiles measured in the steady flows of a 58% suspension, for various rotational velocities ranging from 0.06 to 25rpm.

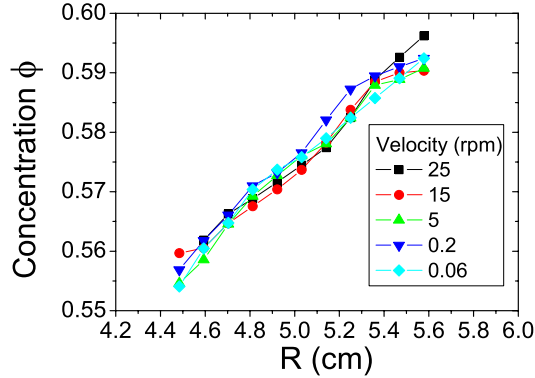

Figure 6: Concentration profiles measured in the gap of the Couette geometry for a suspension of mean volume fraction 58% sheared at various rotational velocities ranging from 0.06 to 25rpm.

We observe in Fig. 6 that the material is inhomogeneous under flow: the concentration is lower near the inner cylinder where the shear rate is higher. The concentration profile does not depend on the inner cylinder velocity $\Omega$: it is the same when all the material is sheared as when the flow is localized; it implies that the shear localization that occurs when decreasing the rotational velocity is not due to changes in the volume fraction. The profile was irreversibly established by the preshear. It is established within a few revolutions (less than 50), and remains stationary over hours (except for very low velocity, where sedimentation is observed). Note however that we did not study systematically the effect of the preshear rotation velocity, and particularly the effect of a preshear by a localized flow: we intend to perform this study later; at this stage, we only observe the same concentration profile for a 9rpm preshear as for a 100rpm preshear.

This is a shear-induced particle migration phenomenon [Acrivos (1995)]: the particles irreversibly migrate towards the low shear zones. However it is much faster than the phenomena usually observed in suspensions of volume fraction up to 55% [Graham *et al.* (1991); Phillips *et al.* (1992); Corbett *et al.* (1995)]: in all these experiments, the authors observe that the migration phenomenon lasts for 800 to 2500 revolutions whereas the phenomenon lasts for less than 50 revolutions in our experiments. This point will be discussed in Sec. IV D, which is devoted to comparison with migration models.

Note however that the material is stationarily inhomogeneous, i.e. all the experiments were performed on the same inhomogeneous structure, for a suspension of a given mean volume fraction.

In Fig. 7, we plot the concentration profiles obtained for several suspensions of mean volume fraction 58%, 59%, and 60%. We observe the same shear induced migration phenomenon in the 3 suspensions, with the same order of magnitude; in the case of the 60% suspension, the concentration seems to saturate at a local 61.5% value near the outer cylinder.

We checked consistency of profiles with mean concentration: for the 58% suspension, a linear fit $\phi = 0.433 + 2.85R$ gives $\bar{\phi} = 0.58$, for the 59% suspension, a linear fit $\phi = 0.44 + 2.84R$ gives $\bar\phi = 0.587$, for the 60% suspension, a linear fit $\phi = 0.48 + 2.4R$ up to 5.6cm and a 61% plateau from 5.6cm to 6cm gives $\bar\phi = 0.6$.

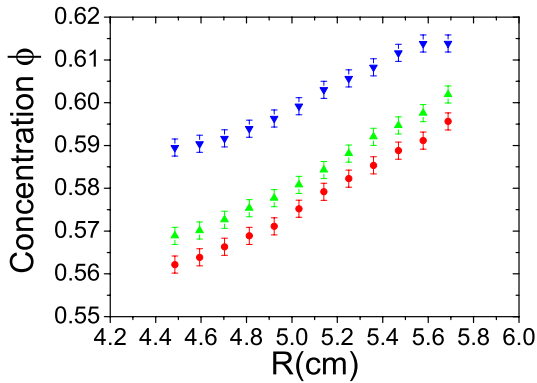

Figure 7: Concentration profiles measured in the gap of the Couette geometry for 3 different sheared suspensions, of mean volume fraction 58% (circles), 59% (triangles), and 60% (down triangles).

## B  Onset of localization

In this section, we study the transient response to a sudden change in rotational velocity. The procedure we apply is the following: (i) we preshear the material at a high rotational velocity $\Omega_0 > \Omega_c$ such that the whole material (in the case of the 58% suspension) is sheared, i.e. in the macro-viscous regime, and then (ii) we instantaneously (duration< 0.1 s) decrease the rotational velocity to a velocity $\Omega_1 < \Omega_c$ in the localized flow regime (i.e. the shear-thinning regime). We then measure the temporal evolution of torque and velocity profiles just after the velocity step. Note however that we cannot directly correlate accurately a MRI measurement and a torque measurement as both are performed on different rheometers (the MRI rheometer does not allow for accurate torque measurements).

We see in Fig. 8a that just after the velocity step the shear torque increases for a deformation roughly independent of velocity, and then reaches an almost stationary value (the slight evolution may be attributed to slight sedimentation at these very low velocities).

This evolution can be correlated with MRI measurements. We see in Fig. 8b and Fig. 9 that during the rapid increase of torque the flow gradually localizes. At short times just after the velocity step, the thickness of the sheared layer is the same as for the preshear velocity, and, when plotted versus the radius $R$, the dimensionless velocity profile $V(R,\Omega_1)/\Omega_1 R_i$ falls along the same curve as the dimensionless velocity profile $V(R,\Omega_0)/\Omega_0 R_i$ during the preshear (Fig. 8b). The thickness of the sheared layer then decreases, as the deformation increases, to its stationary value (Fig. 9), and it results in a stationary torque; the stationary thickness of the sheared layer is lower for lower velocity. The torque increase is thus a macroscopic signature of shear localization.

We also find that for all the velocities $\Omega_1$ studied, at short times just after the velocity step all the dimensionless velocity profiles $V(R,\Omega_1)/\Omega_1 R_i$ fall along the same curve (Fig. 10).

Finally, we see in Fig. 8a and its inset that the relevant parameter that controls localization may be deformation rather than time, as all the torque data roughly follow the same evolution when plotted versus the deformation. The stationary torque is reached for a displacement of the internal cylinder $\Omega_1 t \approx 0.2$rad. In our geometry, this is interpreted as a macroscopic deformation $\gamma_{macro} \approx 0.4$. From local velocity measurements (Fig. 3, 8b) we know that the shear rate at the walls is about 4 times the macroscopic shear rate. Therefore, the local deformation $\gamma_{local}$ at the walls is of order unity.

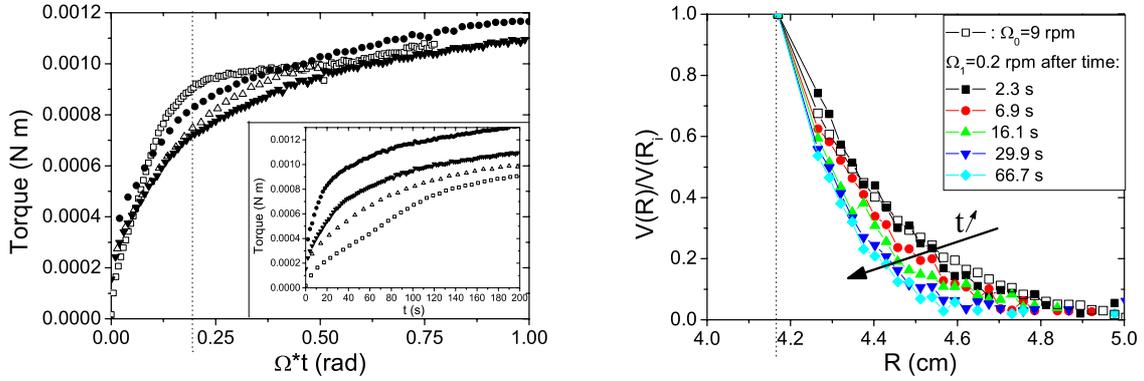

Figure 8: a) Torque vs. deformation angle $\Omega t$ for various velocities $\Omega_1$ just after preshearing the material at $\Omega_0 = 100$rpm ($\Omega_1 = 0.1$rpm (circles), 0.05rpm (down triangles), 0.025rpm (open up triangles), 0.01rpm (open squares)); the inset shows the same data in a torque vs. time plot. b) Reduced velocity $V(R)/V(R_i)$ during a preshear at $\Omega_0 = 9$rpm (open squares) and for various times after the sudden change to velocity $\Omega_1 = 0.2$rpm (for MRI measurements at low velocities, a 100rpm preshear could not be applied). These data were obtained for a 58% suspension.

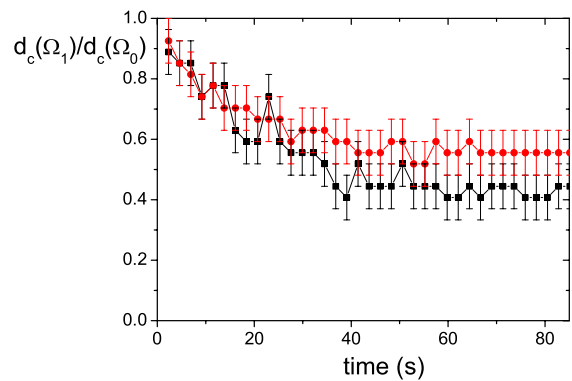

Figure 9: Fraction of material sheared vs. time after the velocity step from a velocity $\Omega_0 = 9$rpm to a velocity $\Omega_1 = 0.2$rpm (circles) and $\Omega_1 = 0.07$rpm (squares). These data are inferred from MRI velocity profile measurements, for a 58% suspension.

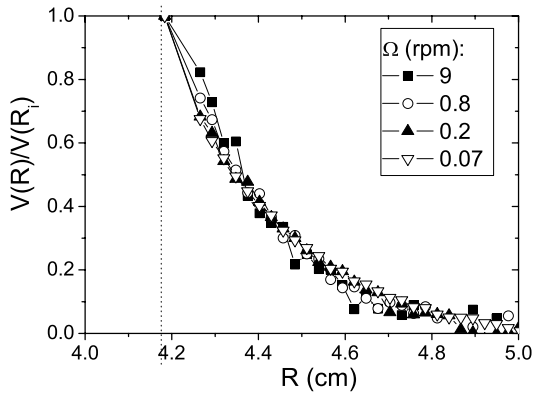

Figure 10: Reduced velocity $V(R)/V(R_i)$ during the preshear at $\Omega_0 = 9$rpm and for various velocities $\Omega_1$ (0.8rpm, 0.2rpm, 0.07rpm), 2 seconds after the sudden change in velocity. These data were obtained for a 58% suspension.

## IV Analysis

### A  Local determination of the constitutive law

Usually, the constitutive law of a material, i.e. a law between the shear stress $\tau$ and the shear rate $\dot\gamma$, may be inferred from a rheometric experiment, which provides stationary torque values $T$ as a function of the rotational velocity $\Omega$. Three important assumptions may then be made:
(i) the shear stress is assumed to be constant along the cylinder's height; this allows to infer the shear stress value from the torque measurement
(ii) the flow is homogeneous in the gap; this allows to infer the shear rate value from the rotational velocity measurement
(iii) the material is homogeneous; in the case of a suspension of mean volume fraction $\bar\phi$, it ensures that the constitutive law found is that of the material at concentration $\phi = \bar\phi$.

Therefore, if we want to build a constitutive law from rheological measurements in a dense suspension, we face two major problems which contradict the points (ii) and (iii): there is shear localization, and the thickness of the sheared layer depends on the rotational velocity for $\Omega < \Omega_c$; the material is inhomogeneous. Note that migration seems to be a quasi instantaneous phenomenon, so that it would not be possible to avoid it by performing measurements at short times after placing only. As a consequence, it seems *a priori* very difficult to infer any information on the behavior of a suspension of a given volume fraction from macroscopic measurements. It casts doubt on the conclusion drawn from previous measurements in the shear thinning regime of these systems: all the interpretations, as regards the nature of dissipation, were based on macroscopic measurements alone.

Nevertheless, two important features will help us to get round these difficulties and infer a constitutive law from the rheometric measurements:
(i) the material is stationarily inhomogeneous, and the concentration profile does not depend on velocity whatever the flow regime, i.e. all the experiments were performed on the same inhomogeneous structure, for a suspension of a given mean volume fraction
(ii) in the velocity step experiments of Sec. III B, the thickness of the sheared layer just after

the abrupt change in velocity is the same for all velocities (Fig. 10).

In Fig. 11, we plot all of the torque values obtained after the velocity step experiments of Sec. III B (see Fig. 8), including the transient values, versus the rotational velocity $\Omega$, for all the velocities $\Omega_1$ studied. We recall that these experiments consisted in decreasing instantaneously the rotational velocity from $\Omega_0 > \Omega_c$ to $\Omega_1$. We observe that the short time torque measurements for all the velocities $\Omega_1$ studied all fall on the same curve ($T = \alpha\Omega$) which is that of a purely viscous material and is the same as in the macro-viscous regime, where all the material is sheared. When $\Omega_1 < \Omega_c$, the torque values then increase with the deformation while shear localization takes place, and the stationary torque values may be fitted to a Bingham law.

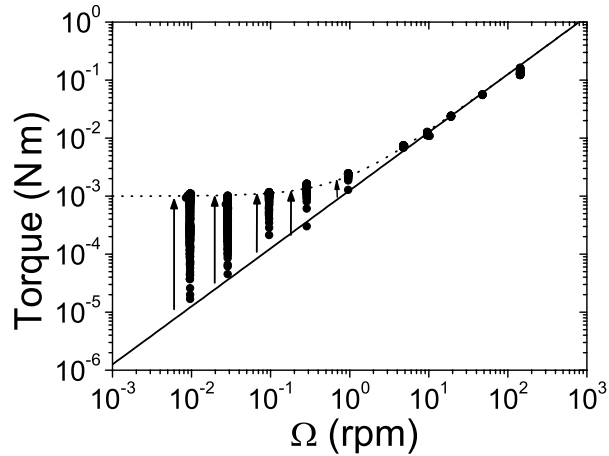

Figure 11: Torque vs. internal cylinder rotational speed for a 58% suspension. The circles were obtained at various rotational velocities $\Omega_1$, ranging from 0.01rpm to 50rpm, just after preshearing the material at $\Omega_0 = 100$rpm before each new velocity during 30s; all the temporal evolution of torque (cf. Fig. 8a) for each velocity $\Omega_1$ is represented, the direction of evolution being indicated by the arrows. The dotted line is a fit of the steady state data (cf. Fig 1) to a Bingham model $T = T_0 + \alpha\Omega$ with $T_0 = 0.001$N m and $\alpha = 0.012$N m/s; the line is a fit to a viscous model $T = \alpha\Omega$ with $\alpha = 0.012$N m/s. The arrows indicate the evolution of torque with the deformation as localization occurs.

The explanation is that at short times, all the material is sheared (Fig. 9): in this case, the conditions are fulfilled for a correct rheometric characterization of a given structure (the stationarily inhomogeneous material) i.e. for a constant sheared layer thickness, and we can conclude from the macroscopic measurements that the inhomogeneous material macroscopic behavior is that of a purely viscous structure, even for very low velocities; there is no observable influence of friction. This purely viscous behavior is observed over 4 decades of velocity (2 decades in the apparent shear-thinning regime, 2 decades in the macro-viscous regime).

We now focus on the local behavior, and show with the help of the velocity profiles that it is a purely viscous behavior. The torque measurements $T$ just after the velocity step provide a macroscopic law $T = \alpha\Omega_1$ for all $\Omega_1$ values. From the reasonable hypothesis of homogeneous stress $\tau_i$ along the internal cylinder, and as $\tau(R) = \tau_i R_i^2/R^2$ in a Couette geometry whatever the constitutive law may be, we get a shear stress distribution

$$\tau(R, \Omega_1) = \frac{\alpha\Omega_1}{2\pi R^2 h} \qquad (6)$$

and thus the local stress at radius $R$ just after the velocity step is simply proportional to the rotational velocity: $\tau(R, \Omega_1) \propto \Omega_1$.

Moreover, for all the velocities in the macro-viscous regime, and for all velocities $\Omega_1$ studied, at short times $t = 0^+$ just after the velocity step, we find that all the dimensionless velocity profiles $V(R, \Omega_1)/V(R_i)$ plotted vs. the radius fall along the same non-Newtonian curve (Fig. 3, 10):

$$V(R, \Omega_1, t = 0^+) = \Omega_1 R_i f(R) \qquad (7)$$

This may read "the local shear rate $\dot{\gamma}(R, \Omega_1, t = 0^+)$ just after the preshear is proportional to the rotational velocity" i.e.

$$\dot{\gamma}(R, \Omega_1, t = 0^+) = \frac{\dot{\gamma}(R, \Omega_0)}{\Omega_0} \Omega_1 \qquad (8)$$

We finally conclude that the local constitutive law reads

$$\tau(R, \Omega) = \eta(R) \dot{\gamma}(R, \Omega) \qquad (9)$$

where $\eta(R)$ depends on the position in the gap, but is everywhere a viscosity independent of the local shear rate. Therefore, the local constitutive law of the flowing material is that of a purely viscous material (we say viscous and not Newtonian as these materials develop normal stresses); there is no observable shear thinning behavior for low shear rate, i.e. the flowing material, in the state prepared by the preshear, does not behave like a dry granular material at low shear rates.

The apparent shear thinning behavior observed in macroscopic experiments may therefore be seen as an artefact induced by shear localization. Actually, we observe that viscous flows at low velocities ($\Omega < \Omega_c$) are not stable, and that the flow localizes. The mechanical origin of localization was evidenced by Huang *et al.* (2005) in a viscosity bifurcation experiment: there is no steady flow for a shear rate below a given critical shear rate $\dot{\gamma}_c$ that depends on the material's properties. When the rotational velocity is lower than $\Omega_c$, the apparent shear rate in the gap is lower than $\dot{\gamma}_c$. Therefore, the flow has to localize so as to ensure that in the flowing material $\dot{\gamma} = \dot{\gamma}_c$. During localization, the shear stress increases while $\dot{\gamma}$ tends to $\dot{\gamma}_c$: if the flowing material constitutive law is that of a purely viscous material, the stationary shear stress at the walls is then approximately $\tau \approx \eta(\phi) \dot{\gamma}_c = \tau_c$. Shear localization thus leads to a shear stress plateau $\tau_c$ at low velocities, i.e. to the apparent Bingham behavior (Fig. 11). This behavior is consistent with what is found in many yield stress fluids [Coussot (2005)], where shear localization [Coussot *et al.* (2002b)] and viscosity bifurcation [Coussot *et al.* (2002a); da Cruz *et al.* (2002)] are observed.

From rheometric experiments performed on a given structure for a constant thickness sheared, and with the help of the velocity profiles measurements, we have thus shown that the local material behavior is a purely viscous behavior. This result was found in the 58% suspension, when all the gap is sheared, i.e. when the material was presheared in the macro-viscous regime. The same has to be shown for the flowing material in the sheared layer, when localization emerges (i.e. when increasing the volume fraction or decreasing the velocity). In the case of the 59 and 60% suspensions, when the material is presheared in the macro-viscous regime, we observe the same features as on Fig. 11; this shows that the flowing behavior of the material in the sheared layer is again a purely viscous behavior, even if a fraction of the material is not sheared in these materials (see Fig. 5). In the case of the apparent shear-thinning behavior, for $\Omega < \Omega_c$, the idea is to perform such velocity step experiments, but now after a preshear at low velocity, in the localized flow regime, so that we start from a localized state and we can characterize the material sheared in a given thickness (smaller than the gap). However, these experiments are hard to perform as there remains only a small layer of the material to be stopped by a decrease in the velocity. In order to study accurately the behavior of the flowing material when shear is localized, we performed another kind of experiments, which show that the flowing material behavior is also a purely viscous behavior when shear is localized; the new method and its results will be presented elsewhere. We also checked consistency of Eq. (9) for all velocities with the help of MRI measurements (see Sec. IV B), i.e. the local viscosity is everywhere independent of the rotational velocity $\Omega_i$ even when the flow is localized ($\Omega_i < \Omega_c$); this shows that the flowing material behavior is not changed when the flow is localized.

Finally, we conclude that the flowing behavior of the material is always a purely viscous behavior, even when the flow is localized, in the range of velocities studied. Note however that we could not study very low velocities, for which we expect the flow to be strongly localized as for dry granular materials. In the next section, we investigate the dependence of the behavior on the volume fraction.

## B  Local viscosity measurements

From rheometric and velocity profiles measurements, we have shown that the local constitutive law of the flowing material is that of a purely viscous material. From concentration profiles measurements, we observe that the material is inhomogeneous. However, we have shown that the material inhomogeneity is stationary and independent of velocity; therefore, it allows to make a change of variable from radius $R$ to concentration $\phi$: all data extracted from measurements at the same radius $R$ in the gap deal with the same material of same volume fraction $\phi$.

Therefore, the shear rate measurements in the gap $\dot{\gamma}(R) = V/R - \partial V/\partial R$, inferred from the velocity profiles $V(R)$, allow, from a unique MRI experiment, to obtain the concentration dependence of viscosity

$$\eta(\phi) = \frac{\tau\big(R(\phi)\big)}{\dot{\gamma}\big(R(\phi)\big)} \qquad (10)$$

A single torque measurement performed on the suspension is also necessary as we need the value of the shear stress $\tau(R_i)$ on the internal cylinder in order to compute the local stress $\tau(R)$ with Eq. (6).

It is important here to recall that the viscosity obtained with Eq. (10) is not just the ratio of shear stress to shear rate that can be computed for any material: it is a constant viscosity (independent of the shear rate) characterizing the purely viscous flow of the material, as was shown in the previous section.

In Fig. 12, we plot the local viscosity data, normalized by the interstitial fluid viscosity, obtained from a single experiment performed on a 59% suspension and compare these data to data obtained from macroscopic torque measurements at various mean volume fractions $\phi$. These macroscopic measurements are obtained under the assumption - false in our experiment, but necessary when a classic macroscopic experiment with no local measurements is performed - that there is a homogeneous Newtonian flow in the gap. We observe that the local measurements may give a viscosity 5 times higher than what would be inferred from a macroscopic experiment; the reason is that the real local shear rate $\dot{\gamma}(R)$ at radius $R$ where the suspension is really 59% is 5 times lower than the mean shear rate $\dot{\gamma}_0$: the hypothesis of a homogeneous Newtonian flow leads to overestimating the shear rate $\dot{\gamma}$ of the 59% suspension by a factor 5.

Actually, as we observe almost instantaneously migration, it seems impossible to perform a good macroscopic characterization of the concentration dependence of viscosity (at least in a Couette geometry). We may then think that all previous measurements performed in a Couette geometry on dense suspensions (55-60%) of non-colloidal particles are erroneous and underestimate the concentration dependence of viscosity; this underestimation could be avoided for less dense suspensions as migration may be much slower in this case (see Sec. IV D).

Note however that when plotting viscosity versus concentration, there is an important assumption that the viscosity depends only on the concentration. We know that this is not true as the microstructure of the material at a given volume fraction may evolve under shear, e.g.

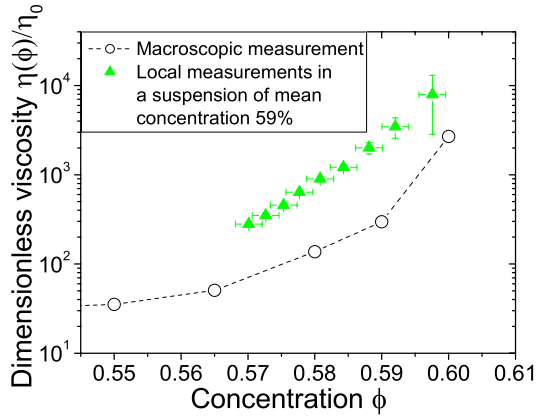

Figure 12: Local viscosity measurements in a 59% suspension (triangles), and macroscopic viscosity measurements for 55 to 60% suspensions (open circles). The viscosity values are normalized by the interstitial fluid viscosity $\eta_0$.

to a more ordered state [Gondret and Petit (1996); Völtz *et al.* (2002)], of smaller viscosity. In our experiments, we did not observe this evolution at the time scale of the experiments (the velocity profiles do not vary with time), but there still is a possibility that the microstructure of the locally 56% suspension in a suspension of mean volume fraction 59% is different from the microstructure of the locally 56% suspension in a suspension of mean volume fraction 56%. However, at this stage we cannot observe such effects and our only control parameter is the volume fraction.

Here, with this new method, from a single experiment on a given suspension, we are able to obtain the concentration dependence of viscosity in a given range of concentrations (in the case of the 59% suspension, from 57 to 60%), which is obtained from the real local shear rate and the real local concentration. With regard to curvilinear flows, it is only possible in a Couette geometry as only in this geometry do we know the local value of stress, independently of the constitutive law; it would also be possible in pipe flows, where we also know the local value of stress (this property was used by Powell *et al.* (1994) to measure the viscosity of a polymeric fluid). Our experiments may thus provide a fair experimental basis for flow predictions and for comparison with theoretical models.

It is important to evaluate how these local measurements may be affected by measurement problems, i.e. wall slip and a slight sedimentation. We can never be sure that there is no slip at the walls in our experiments since we cannot make accurate measurements of the material's velocity near the inner cylinder; we tried to avoid this phenomenon through the use of sandpaper, and the velocity profiles we measure show that if there is wall slip, it is small. However, wall slip would not affect our local determination of the constitutive law nor our measurements of viscosity: the validity of the measurements of the local values of stress, shear rate, and concentration does not depend on what happens at the walls, i.e. we are able to measure locally the viscosity whether there is wall slip or not. Another problem is that the inner cylinder is about 4cm from the bottom, and that there still is a competition between resuspension and sedimentation. This leads to two problems: the resting suspension at the bottom may act as a particle reservoir (as in Gadala-Maria and Acrivos (1980) experiments), i.e. the mean volume fraction in the gap is likely to be different from the mean volume fraction of the suspension; there is a slight vertical inhomogeneity (in the upper part of the gap, the suspension is slightly less dense). Once again, this does not affect the validity of the measurements of the local values of shear rate and concentration. However, stress measurements are affected: the contribution of the bottom of the material to the torque may be different for different suspensions; and we cannot know exactly the stress value at the inner cylinder for our slice measurement if there is a vertical inhomogeneity. Nevertheless, the stress distribution in the gap is not affected: it is still given by $\tau(R) = \tau(R_i)R_i^2/R^2$; as a consequence, the ratio of local viscosities measured at different radiuses $R_1$ and $R_2$, $\eta(R_1)/\eta(R_2)$, is independent of what is measured at the walls. To sum up, our experiments provide a perfect measurement of the evolution of the local viscosity with the concentration, whatever the rheometric measurements problems may be; the absolute value of viscosity depends on the torque measurement and may thus be affected by measurements problems. We plan to improve the measurement geometry and to build a torque probe for our home-made NMR-compliant rheometer.

In the following, we trust the absolute value of the local viscosities inferred from MRI and torque measurements on the 59 and 60% suspensions, for which no vertical inhomogeneity is observed. In the case of the 56.5 and 58% suspensions, we infer the evolution of local viscosity with the concentration from the MRI measurements, and the absolute value of viscosity is obtained by ensuring consistency between two local measurements at the same local concentration in two different suspensions. We can now combine the local viscosity values obtained in a 59% suspension with values obtained in the other suspensions, in order to obtain the local concentration dependence of viscosity for a wider range of concentrations.

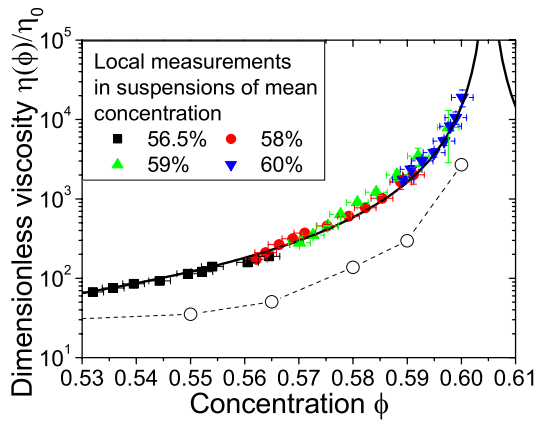

Figure 13: Local and macroscopic viscosity measurements on suspensions of various mean concentration ranging from 55% to 60%. The line is a fit to a Krieger-Dougherty law $\eta(\phi) = \eta_0(1 - \phi/0.605)^{-2}$.

We see in Fig. 13 that these first local viscosity measurements are well fitted to a Krieger-Dougherty law with a maximum packing fraction $\phi_m = 0.605$ and an exponent $n = 2$

$$\eta(\phi) = \eta_0 \frac{1}{\left(1 - \frac{\phi}{0.605}\right)^2} \qquad (11)$$

## C  Consequences of the existence of a maximum packing fraction

We found that the flowing material constitutive law is that of a purely viscous material and that the viscosity diverges at a value of the packing fraction $\phi_m = 0.605$: this implies that there may be no flow for $\phi > \phi_m$. This explains why in the macro-viscous regime of a 59% suspension, we observed that the gap cannot be fully sheared (Fig. 4): there still is a 3-4mm region near the outer cylinder where the material is not sheared; when the rotational velocity increases, the thickness of the sheared layer does not increase anymore for $\Omega_i > \Omega_c$. This is consistent with the concentration profile we observe in Fig. 7: even if we have no measurement in this zone, from extrapolation of the concentration profile, we expect $\phi > \phi_m$ for $R > 5.7$cm. We also observed in Fig. 5 that in the macro-viscous regime of a 60% suspension, the material is sheared up to around 5.1cm: this is also in agreement with our interpretation since we observe on the concentration profile (Fig. 7) that for a radius $R >5.2$cm, we have $\phi > \phi_m = 0.605$. In the case of a suspension of mean volume fraction 58%, we have seen that in the macro-viscous regime all the gap is sheared: this is again consistent with the concentration profile as we observe that $\phi < \phi_m$ everywhere in the gap.

We conclude that for a highly concentrated suspension, shear-induced migration prevents the material from being fully sheared in the macro-viscous regime: the particles migrate out of the high shear zone and cause the formation of a zone of packing fraction higher than the maximum packing fraction, which cannot be sheared in the macro-viscous regime (for a very high velocity, a turbulence or a collisional regime may emerge, for which the situation may be different). Note however than if we could observe the onset of migration, we would expect all the gap to be sheared at the beginning of shear for the homogeneous material. This feature may be used for the measurement of the maximum packing fraction: $\phi_m$ is the concentration where the material stops flowing in the macro-viscous regime; this value may be more accurate than the one inferred from a fit of the viscosity data to a given viscosity law, which may moreover depend on the viscosity law that is chosen.

The macro-viscous regime of dense suspensions, which is characterized by a linear increase of torque with the rotational velocity, then starts when the sheared region has reached a region where $\phi = \phi_m$ (in the case of the 59 and 60% suspensions) or the outer cylinder if $\phi < \phi_m$ everywhere in the gap (in the case of the 58% suspension): the thickness of the sheared layer cannot increase anymore, and now the torque has to increase linearly with the rotational velocity since the local shear rate now increases linearly with the velocity and since the flowing material constitutive law is that of a purely viscous material. A consequence of this phenomenon is that it would be possible to perform rheometric experiments in a suspension of mean concentration $\bar\phi > \phi_m$, and to observe a macro-viscous regime resulting from the shear of a layer of packing fraction lower than $\phi_m$ near the inner cylinder; the wrong conclusion drawn from such macroscopic experiments, if there is no local observation, would then be that the maximum packing fraction is higher than the real $\phi_m$ of the material

Note that this localization, which is simply due to the existence a maximum packing fraction above which there can be no viscous flow, should not be mistaken for the localization which occurs when decreasing the velocity (see Sec. III A); in the latter case, localization occurs on the same structure, without any change in the volume fraction profile, therefore for a volume fraction lower than the maximum packing fraction, and may be attributed to the existence of a critical shear rate $\dot\gamma_c$ below which no steady flow exists.

## D  Comparison with migration models

From our experimental results, two results can be compared to migration models: the duration of the migration phenomenon, and the stationary concentration profiles.

### 1  Migration dynamics

In the diffusive model of Leighton and Acrivos (1987b) and Phillips *et al.* (1992), the particles are submitted to a shear-induced diffusion, and the diffusion coefficient $D$ may be written as [Leighton and Acrivos (1987a); Acrivos (1995)]:

$$D = \bar{D}(\phi)\dot\gamma a^2 \qquad (12)$$

where $\phi$ is the volume fraction, $\dot\gamma$ the shear rate, $a$ the particle size, and $\bar{D}$ is the dimensionless diffusion coefficient, whose dependence on $\phi$ may theoretically be $\bar{D}(\phi) \propto \phi^2$; $\bar{D}(\phi)$ may actually be a tensor [Breedveld *et al.* (2002)]. The gradients in shear rate in a Couette geometry then generate a particle flux towards the outer cylinder, which is counterbalanced by a particle flux due to viscosity gradients. As qualitatively confirmed experimentally by Abbott *et al.* (1991) and Corbett *et al.* (1995), one would then expect the migration phenomenon to last for a number of revolutions

$$N_{migr} \propto \frac{(R_e - R_i)^3}{\bar{R}a^2\phi^2} \quad (13)$$

until the stationary profile is established, where $R_e$ and $R_i$ are respectively the external and internal radius, and $\bar{R} = (R_e + R_i)/2$.

We evaluate the expected $N_{migr}$ in our experiment from experimental results from literature: Phillips *et al.* (1992) find a steady state at $N_{migr} = 800$ revs for $675\mu$m particles at mean volume fraction $\bar{\phi} = 0.55$ in a Couette geometry of radiuses $R_e = 2.38$cm and $R_i = 0.64$cm; Corbett *et al.* (1995) find $N_{migr} = 2000$ revs for $140\mu$m particles, at $\bar{\phi} = 0.4$, with $R_e = 1.9$cm and $R_i = 0.95$cm; Graham *et al.* (1991) find $N_{migr} = 2500$ revs for $600\mu$m particles at $\bar{\phi} = 0.5$, with $R_e = 2.38$cm and $R_i = 0.48$cm. These values are roughly consistent with Eq. (13), and imply an expected value of $N_{migr} \approx 1000$ revs in our experiment.

This is much more than we could observe. We never observed transient concentration profiles. As a measurement lasts for 3 minutes, from the preshear velocities imposed and from the velocities studied, we deduce that migration occurs during the first 50 revs, which is much lower than the expected 1000 revs. An indirect indication is also given by the velocity profiles: the velocity profiles we measure are stationary in less than 10 turns i.e. migration may occur during these 10 turns. We thus conclude that if a diffusive process is involved, it is much faster (at least 20 times) than the one observed in previous studies.

One could then put forward a strong dependence of the dimensionless diffusion coefficient $\bar{D}(\phi)$ on concentration (i.e. much stronger than the $\phi^2$ dependence). However, there is no clear experimental evidence in particle diffusion measurements, that there is any departure from the $\phi^2$ dependence [Leighton and Acrivos (1987a); Breedveld *et al.* (1998, 2002)]. Nevertheless, Tetlow *et al.* (1998) found that in order to fit migration experimental data to the shear induced migration model, $\bar{D}(\phi)/\phi^2$ must depend, e.g. linearly, on the volume fraction; it seems to be the only experimental result suggesting this dependence: the experiments of Phillips *et al.* (1992), Graham *et al.* (1991), and Corbett *et al.* (1995) are consistent with each other, for concentrations ranging from 40% to 55%, evidencing a rather smooth dependence of the dimensionless diffusion coefficients on the packing fraction in this range; Leighton and Acrivos (1987b) could also fit their rheometric data to the shear induced migration model, for concentrations ranging from 40% to 50%, with a $\phi^2$ dependence of the dimensionless diffusion coefficients on concentration. The dimensionless diffusion coefficients would thus have to increase a lot above a 55% concentration, so as to explain the discrepancy between these experiments and ours.

However, as far as we know, there are no accurate measurements of the diffusion coefficients for 55 to 60% suspensions. Moreover, as we will see below, from concentration profiles measurements we can only obtain the relative importance of the dimensionless diffusion coefficients associated with particles collisions and gradients in the relative viscosity, therefore we cannot conclude. We intend to study the onset of migration in a next study; this will imply to work at very low velocities.

## 2 Stationary concentration profiles predictions

The next thing we can do is compare the stationary concentration profiles we observed to the predictions of migration models. Here, we study the predictions of two models: the diffusive model of Phillips *et al.* (1992) and Tetlow *et al.* (1998) with and without any dependence of the dimensionless diffusion coefficients on concentration, and the Morris and Boulay (1999) model, where particle fluxes are generated by gradients in normal stresses.

In these previous studies, the authors assumed a divergence of viscosity at a maximum packing fraction $\phi_m = 0.68$, and, in the case of Phillips *et al.* (1992) and Tetlow *et al.* (1998), a Krieger-Dougherty dependence with a divergence exponent $n = 1.82$. Here we do not need to make any assumption as we measured the local concentration dependence of viscosity (Sec. IV B). However, we find a concentration dependence which is different from that assumed in the previous studies. That is why in the following, we will plot and compare the prediction of the models based on the locally measured viscosity with the predictions based on the wrong assumption, in order to check consistency of our results with the results by other authors.

In the Phillips *et al.* (1992) model, based on Leighton and Acrivos (1987b) model, there are two particle fluxes: a first one $N_c$ due to gradients in collision frequency

$$N_c = -K_c \phi a^2 \nabla(\dot{\gamma}\phi) \qquad (14)$$

and a second one due to viscosity gradients

$$N_\mu = -K_\mu a^2 \frac{\dot{\gamma}\phi^2}{\eta(\phi)} \nabla(\eta(\phi)) \qquad (15)$$

The stationary profile then results from competition between both fluxes. The dimensionless diffusion coefficients $K_c$ and $K_\mu$ are first considered as constant in the following (the $\phi^2$ term is explicitly written in Eq. (14,15)), their dependence on $\phi$ will then be considered.

The Phillips *et al.* model, for a Krieger-Dougherty viscosity $\eta(\phi) = \eta_0(1 - \phi/\phi_m)^{-n}$ predicts a steady state:

$$\frac{\phi(R)}{\phi(R_i)} = \left(\frac{R}{R_i}\right)^2 \left(\frac{1 - \phi(R_i)/\phi_m}{1 - \phi(R)/\phi_m}\right)^{n*(1-K_\mu/K_c)} \qquad (16)$$

so that there is only one fit parameter if the viscosity is known, which is the ratio of the dimensionless diffusion constants $K_\mu/K_c$.

We compare the predictions of the steady state of the Phillips migration model with the experimental data in Fig. 14.

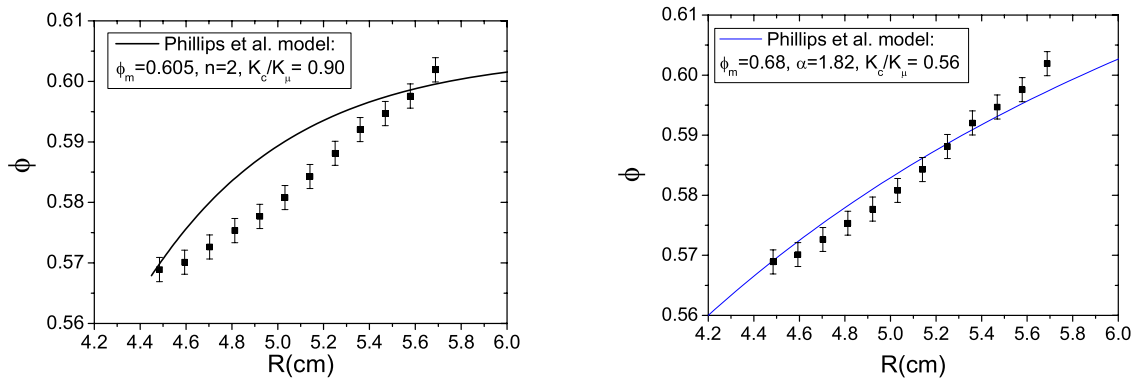

Figure 14: Comparison of concentration profile obtained experimentally for a 59% suspension with the predictions of the Phillips *et al.* model with two expressions of viscosity: a) with the local dependence we measured ($\phi_m = 0.605$, $n = 2$); b) with the dependence postulated by Phillips *et al.*, obtained from macroscopic experiments ($\phi_m = 0.68$, $n = 1.82$).

We see that the predictions of the model may agree with our experimental data if we pre-suppose, as Phillips *et al.*, a Krieger-Dougherty dependence of viscosity with maximum packing fraction $\phi_m = 0.68$, and a divergence exponent $n = 1.82$ (Fig. 14b). However, we see that this is not consistent with our local viscosity measurements inferred from concentration and velocity profiles. A fit based on these last measurements may be trusted, rather than a model-dependent fit. If we try to fit the concentration profiles to the model with the local concentration dependence of viscosity (i.e. with $\phi_m = 0.605$), we find a rather important discrepancy between the experimental data and this model (Fig. 14a). The qualitative origin of this discrepancy, as observed on Fig. 14a, is that the model predicts that $\phi$ tends asymptotically to $\phi_m$ as $R$ increases, whereas experimentally we observe that $\phi$ seems to tend to $\phi_m$ for a finite value of $R$ (as is the case for the 59 and 60% suspensions); this new experimental feature may be a test for any migration model.

It was pointed out by Tetlow *et al.* (1998) that the ratio of the dimensionless diffusion coefficients $K_\mu/K_c$ may depend on the packing fraction $\phi$. Assuming a linear dependence $K_\mu/K_c(\phi) = c\phi + b$, the Phillips *et al.* model, for a Krieger-Dougherty viscosity $\eta(\phi) = \eta_0(1 - \phi/\phi_m)^{-n}$, then gives a steady state:

$$\frac{\phi(R)}{\phi(R_i)} = \left(\frac{R}{R_i}\right)^2 \left[\frac{c\phi(R_i) + b}{c\phi + b}\right]^{n/(c\phi_m + b)} \left(\frac{1 - \phi(R_i)/\phi_m}{1 - \phi(R)/\phi_m}\right)^{n*(1 - 1/(c\phi_m + b))} \quad (17)$$

Note however that by adding this dependence on concentration, the Phillips *et al.* model loses its simple physical justification; and now this is a 2 parameter model. We compare the predictions of this model with our data in Fig. 15. We now observe that the data are well fitted to the model with $K_\mu/K_c = 7.3\phi - 3.4$, if we impose $\phi_m = 0.605$ as we observed experimentally (Fig.15a). This gives a higher dependence of $K_\mu/K_c$ on the concentration $\phi$ than that observed by Tetlow *et al.* (1998) who found $c \approx 1.5$. However, Tetlow *et al.* (1998) had postulated a value of $\phi_m = 0.68$; therefore, in order to compare our data to theirs, we also fit the concentration profiles to the model with $\phi_m = 0.68$ and we impose $c = 1.5$ (see Fig.15b): the data can also be fitted, so that our data are consistent with those of Tetlow *et al.* However, as we measured the local viscosity and found $\phi_m = 0.605$, if this model is valid, the value $c = 7.3$ may be trusted rather than $c = 1.5$, which was based on erroneous postulation of the concentration dependence of viscosity, and may be a fair basis for comparison with direct measurements of shear induced particle diffusion.

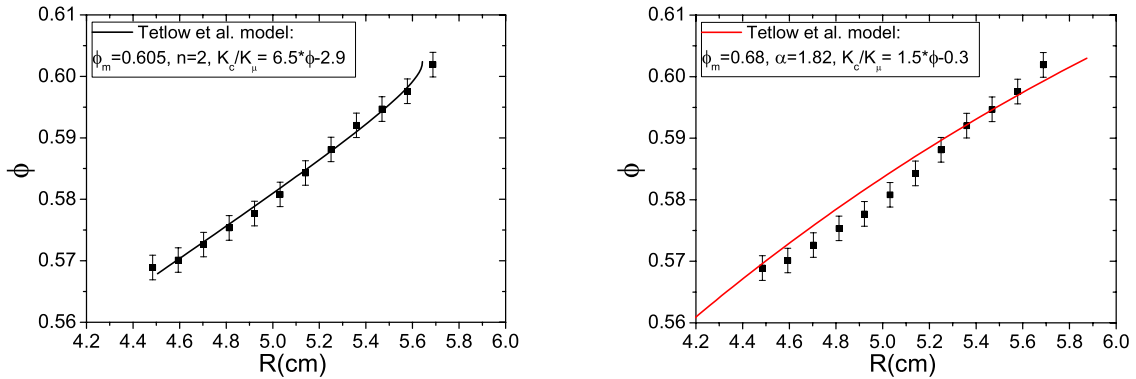

Figure 15: Comparison of the concentration profile obtained experimentally for a 59% suspension with the predictions of the Tetlow *et al.* model with two expressions of viscosity: a) with the local dependence we measured ($\phi_m = 0.605$, $n = 2$); b) with the dependence postulated by Tetlow *et al.*, obtained from macroscopic experiments ($\phi_m = 0.68$, $n = 1.82$).

As an alternative to the diffusive models, one may consider the role of normal stresses [Morris and Boulay (1999)]. In these models, the particle fluxes counterbalance the gradients in normal stresses. In order to predict the stationary concentration profile, one may thus know the value of shear-induced normal stresses which may be written, in tensorial form, as:

$$\Sigma_n = -\eta_n(\phi)\dot{\gamma} \begin{pmatrix} 1 & 0 & 0 \\ 0 & \lambda_1 & 0 \\ 0 & 0 & \lambda_2 \end{pmatrix} \quad (18)$$

As we did not measure these normal stresses, we may assume the same dependence on concentration as Morris and Boulay (1999):

$$\eta_n(\phi) = \eta_0 K_n \frac{(\phi/\phi_m)^2}{(1-\phi/\phi_m)^2} \quad (19)$$

with $K_n = 0.75$. Again, we write a Krieger-Dougherty shear viscosity $\eta_s(\phi) = \eta_0(1-\phi/\phi_m)^{-n}$. The steady state is then given by [Morris and Boulay (1999)]:

$$K_n \frac{(\phi/\phi_m)^2}{(1-\phi/\phi_m)^2}(1-\phi/\phi_m)^n = A_2 R^{(1+\lambda_2)/\lambda_2} \quad (20)$$

and $\lambda_2$ is the only fit parameter ($A_2$ is determined by requiring that the mean volume fraction found with this expression is that of the suspension).

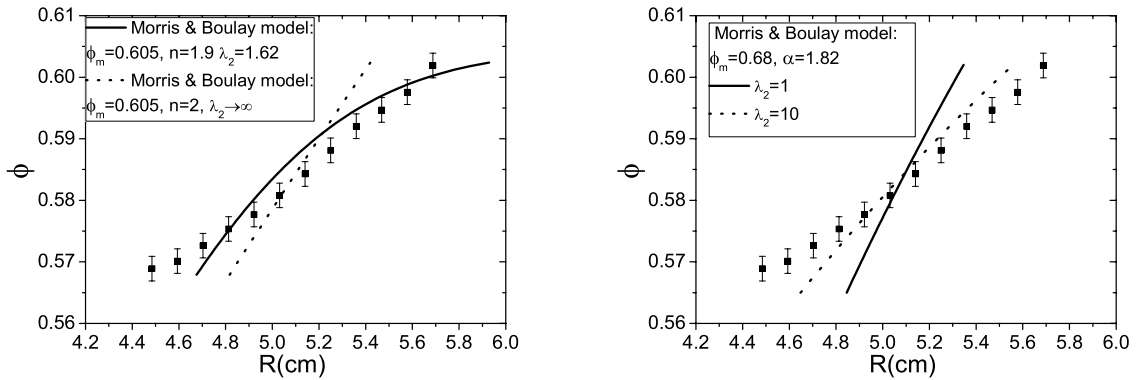

Figure 16: Comparison of the concentration profile obtained experimentally for a 59% suspension with the predictions of the Morris and Boulay model with two expressions of viscosity: a) with the local dependence we measured ($\phi_m = 0.605$, $n = 2$ and $n = 1.9$); b) with $\phi_m = 0.68$ as postulated by Morris and Boulay, and $n = 1.82$.

Eq. (20) is very sensitive to the value of $n$: a value of $n = 2$ is particular since in this case $R$ does not diverge as $\phi$ tends to $\phi_m$; $n > 2$ would require $-1 < \lambda_2 < 0$ in order to ensure that $R$ increases when $\phi$ tends to $\phi_m$; that is why we fit our data to eq. (20) with $n = 2$ and a close value $n = 1.9$ (for $\phi_m = 0.605$ as experimentally measured). We find that the model with $n = 2$ does not fit to our data (and the best fit is for $\lambda_2 \to \infty$), and that the agreement is not perfect if we set $n = 1.9$ (Fig.16a). This imperfection may be attributed to the same qualitative trend as in the Phillips *et al.* diffusive model: the model predicts that $\phi$ tends asymptotically to $\phi_m$ as $R$ increases (for $n < 2$), whereas experimentally we observe that $\phi$ seems to tend to $\phi_m$ for a finite value of $R$. Finally, as for the Phillips *et al.* model, a rough agreement could be found if we erroneously postulated $\phi_m = 0.68$ as was done by Morris and Boulay (1999) in order to compare their model to the Phillips *et al.* model (Fig. 16b), but with a rather high value of $\lambda_2$ (we find $\lambda_2 \approx 10$ whereas Morris and Boulay (1999) found $\lambda_2 \approx 1$). Note however that: (i) Morris and Boulay considered a different expression of viscosity; (ii) Eq. (19) for the normal viscosity $\eta_n(\phi)$ is not supported by any experimental data: it would be important to provide a local measurement of $\eta_n(\phi)$ in order to make a fair comparison of our experimental data with this model; (iii) if there is any theoretical reason that $\eta_n(\phi)/\eta_s(\phi)$ tends to a constant as $\phi \to \phi_m$, then this model ensures that $\phi$ tends to $\phi_m$ for a finite value of $R$ as observed experimentally.

As a summary, we observed a very fast migration dynamics which is in contradiction with diffusive models if we try to check consistency with the dynamics at lower concentrations (40-55%). Moreover, we find a discrepancy between our data and the Phillips *et al.* and the Morris and Boulay models, if we consider the local viscosity we measured: an agreement with other data in literature may be partly attributed to the erroneous postulation that $\phi_m = 0.68$. An agreement may be found with the Tetlow *et al.* model, but this model now lacks simple physical meaning as the dimensionless diffusion coefficients dependence on packing fraction is added *ad hoc*, and has a supplementary fit parameter.

# V   Synthesis: constitutive law and open problems

In this section, we propose a constitutive law for dense non-colloidal suspensions, try to relate it to the physical properties of the material, and show how it accounts for the macroscopic observations.

The main observations we have to understand are: (i) the macroscopic rheological measurements exhibit an apparent yield stress, (ii) at controlled velocity, for low velocities below a critical velocity $\Omega_c$, the shear flow is localized and the shear stress value is roughly constant (this is the apparent yield stress); above $\Omega_c$ all the gap is sheared and the shear stress is proportional to the rotational velocity, (iii) at controlled stress, this system exhibits a viscosity bifurcation [Huang *et al.* (2005)] i.e. below a critical stress $\tau_c$ (the apparent yield stress) there is no flow, and above $\tau_c$ the rotational velocity is higher than a critical value $\Omega_c$ (this value is the same as that below which shear localization is observed at controlled velocity), (iv) this system exhibits normal stresses [Zarraga *et al.* (2000)] and normal stresses are proportional to the shear stress in both regimes [Prasad and Kytömaa (1995); Huang *et al.* (2005)].

The very simple law we propose, which is reminiscent of what happens in pastes [Coussot (2005)], is the following: if the local shear rate $\dot\gamma$ is higher than a critical shear rate $\dot\gamma_c(\phi)$ which may *a priori* depend on the volume fraction $\phi$, then the material flows and its shear stress is purely viscous

$$\dot\gamma > \dot\gamma_c(\phi) \; : \; \tau = \eta(\phi)\dot\gamma \tag{21}$$

with

$$\eta(\phi) = \eta_0(1 - \phi/\phi_m)^{-n} \tag{22}$$

where the maximum packing fraction is $\phi_m = 0.605$ and the divergence exponent is $n = 2$; above the maximum packing fraction $\phi_m$, no flow is allowed (see Sec. IV C).
The normal stress follows a law

$$\sigma_N = \beta(\phi)\dot\gamma \tag{23}$$

An experimental determination of $\beta(\phi)$ may be found in [Zarraga *et al.* (2000)]; however, we think that the knowledge of the concentration dependence of these measurements can probably be improved by local measurements as those we performed for viscosity in Sec. IV B.
If $\dot\gamma < \dot\gamma_c(\phi)$ then the material stops flowing after a deformation of order 1 (see Sec. III B).

We showed in Sec. IV A and Sec. IV B that the flowing material behavior is a purely viscous behavior, and that the viscosity follows Eq. (22); we shall not discuss these points again. In the following, we discuss the criterion that controls the jamming transition.

In the unsheared zone, the material is jammed. There may then be a contact network in this part of the material in order to sustain a yield stress without flowing, since there may be no other interactions as hydrodynamic interactions and direct contacts. As a consequence, we would expect the criterion for starting and stopping (as localization is perfectly reversible) the flow to be a frictional Coulomb criterion: $\tau/\sigma_N \leq \mu_s(\phi)$ in the jammed material (for a discussion on the $\phi$ dependence of the friction coefficient $\mu_s$ near the jamming transition, see da Cruz *et al.* (2005)). At the interface between the jammed material and the flowing material, at radius $R_c(\Omega)$, the Coulomb criterion would then be met. On the flowing material side, the shear and normal stresses values are known: $\tau(R_c) = \eta(\phi(R_c))\dot\gamma(R_c)$ and $\sigma_N(R_c) = \beta(\phi(R_c))\dot\gamma(R_c)$; the Coulomb criterion would then read: $\eta(\phi(R_c))\dot\gamma(R_c) = \mu_s(\phi(R_c))\,\beta(\phi(R_c))\dot\gamma(R_c)$, and this value should be the apparent yield stress measured in the rheometric experiments. Finally, the Coulomb criterion would read

$$\eta(\phi(R_c)) = \mu_s(\phi(R_c))\,\beta(\phi(R_c)) \qquad (24)$$

which is a criterion independent of the shear rate, i.e. Eq. (24) would define a static interface between the jammed material and the flowing material. As a consequence, we cannot explain the evolution with the shear rate of the position of the interface between the sheared and the unsheared material with a Coulomb criterion. Let us recall here that the concentration profile is independent of the rotational velocity (see Sec. III A 3) so that localization is not due to changes in the volume fraction when the velocity decreases.

Thus another jamming criterion than the Coulomb criterion must be proposed. The existence of a critical shear rate $\dot\gamma_c(\phi)$ below which no steady flow exists seems to be a fair and necessary criterion. It accounts for the existence of an apparent yield stress and for the viscosity bifurcation experiments of Huang *et al.* (2005): when a shear stress $\tau(R_i)$ is imposed at the inner cylinder, the shear rate at the inner cylinder in a steady state is: $\dot\gamma(R_i) = \tau(R_i)/\eta(\phi(R_i))$; when the shear stress is decreased below the critical shear stress $\tau_c = \eta(\phi(R_i))\,\dot\gamma_c(\phi(R_i))$, then no steady flow is allowed (as it would imply $\dot\gamma(R_i) < \dot\gamma_c$), and the macroscopic flow gradually stops. It also accounts for shear localization: when the rotational velocity decreases, and is such that the mean shear rate is lower than $\dot\gamma_c$, the thickness of the sheared layer has to decrease so as to ensure $\dot\gamma(R) > \dot\gamma_c(\phi(R))$ at each radius $R$ in the sheared layer.

We may now understand the macroscopic behavior observed in the rheometric experiments, i.e. a torque plateau followed by a linear regime, and a torque proportional to normal forces in both regimes. At low velocities, in the apparent shear-thinning regime, there is a coexistence between a sheared and an unsheared region so as to ensure $\dot\gamma > \dot\gamma_c$ in the sheared region. The limit between both regions is in $R = R_c(\Omega)$, such that $\dot\gamma(R_c) = \dot\gamma_c(\phi(R_c))$; the stress at the inner cylinder, in $R = R_i$, is then

$$\tau(R_i) = \tau(R_c)R_c^2/R_i^2 = \eta(\phi(R_c))\dot\gamma_c(\phi(R_c))R_c^2/R_i^2 \qquad (25)$$

i.e. the coexistence between a sheared and an unsheared region results in a roughly constant torque and in an apparent yield stress

$$\tau_c = \eta(\phi(R_i))\dot\gamma_c(\phi(R_i)) \qquad (26)$$

At low velocity, a small increase in the rotational velocity then implies a slight increase in the torque for two reasons: (i) because $R_c$ increases; (ii) because the material is inhomogeneous (i.e. $\phi(R_c)$ increases). When the sheared region has reached a region where $\phi = \phi_m$ (in the case of the 59 and 60% suspensions) or the outer cylinder if $\phi < \phi_m$ everywhere in the gap (in the case of the 58% suspension), the thickness of the sheared layer cannot increase anymore, and now the torque has to increase linearly with the rotational velocity since the local shear rate now increases linearly with the velocity and since the flowing material constitutive law is that of a purely viscous material. In both regimes, the shear stress is generated by viscous dissipation, i.e. $\tau(R_i) = \eta(\phi(R_i))\dot\gamma(R_i)$ whatever the apparent flow regime may be; the material then also develops normal stresses $\sigma_N(R_i) = \beta(\phi(R_i))\dot\gamma(R_i)$. Therefore, at the walls, we always have a proportionality between torque and normal forces as

$$\tau(R_i)/\sigma_N(R_i) = \eta(\phi(R_i))/\beta(\phi(R_i)) \qquad (27)$$

with the same value in both regimes, as observed experimentally.

We finally comment on the physical origin of this behavior. In the unsheared zone, the material is jammed. Therefore, there may be a contact network in this part of the material in order to sustain a yield stress without flowing, since there may be no other interactions as hydrodynamic interactions and direct contacts. On the other hand, in the flowing material zone, we have shown that the local behavior is a purely viscous behavior even at very low shear rates: now there may be no direct frictional contacts, and the only dissipative interactions are hydrodynamic interactions. A natural interpretation of localization is that at the critical shear rate $\dot\gamma_c$ there is a structural change in grain configuration from a flowing state with no direct contacts to a contact network in a jammed state (near the outer cylinder). We have observed in Sec. III B that the relevant parameter that controls localization is deformation rather than time: localization occurs for a deformation of order unity. The value $\gamma_{local} \approx 1$ is consistent with a change in configuration at the grain scale as it corresponds to a displacement of one grain with respect to another grain.

The existence of a critical shear rate remains however to be understood. A possibility is that there is competition between viscous normal forces and buoyant forces, as suggested by Ancey and Coussot (1999). In this case, one has to compare the vertical scale of normal stresses in the contact network, $\Delta\rho\phi gh$ to the viscous normal force $\beta(\phi)\dot\gamma$; one thus gets $\dot\gamma_c(\phi) = \Delta\rho\phi gh/\beta(\phi)$. This proportionality with $h$ is supported by the macroscopic data of Ancey and Coussot (1999) for glass beads, while an invert proportionality with the interstitial fluid viscosity was reported by Huang *et al.* (2005). However, as our particles are close to be neutrally buoyant, it is likely that another scale of normal forces in the contact network should be introduced.

Note that we do not deny the existence of frictional flows in dense suspensions, but rather their identification to the shear plateau on the macroscopic rheometric measurements. Indeed, frictional flows in granular materials are strongly localized: the sheared layer is of about 5 to 10 grains large [GDR Midi (2004)]. At very low velocities, the flow in dense suspensions will be strongly localized, and a decrease in velocity will not produce any additional localization. We then expect the material to change its structure and the flow to become frictional (if the local concentration at the inner cylinder is large enough). As a consequence, there may be two critical shear rates: a first one, $\dot\gamma_{c1}$, for a change from a frictional flow to a viscous flow, a second one, $\dot\gamma_{c2}$, for the onset of localization of the shear flow of a viscous dense suspension. As far as we understand it, $\dot\gamma_{c1} < \dot\gamma_{c2}$. In our experiments, we would have to go below $\Omega = 0.01$rpm in order to observe such strongly localized flows, so we could not study this regime. A possibility is to study less viscous interstitial fluids since they exhibit higher $\dot\gamma_{c2}$ [Huang *et al.* (2005)]. Ancey and Coussot (1999) were aware of the existence of a transition zone between a frictional flow and a viscous flow, although they could not identify nor characterize it; in their work, both regimes (localized viscous flow and frictional flow) may be responsible for the shear plateau which is observed for a large range of rotational velocities.

To complete the picture, we would also need to have a migration model as shearing a dense suspension may always result in an inhomogeneous state. However, at this stage, we have seen that migration is far from being understood.

# VI  Conclusion

We have studied the flowing behavior of dense suspensions of non-colloidal particles, by coupling local velocity and concentration measurements through MRI techniques, and macroscopic rheometric experiments in a Couette geometry.

We showed that the flow is localized at low velocities and also at high concentrations. We also showed that the material is inhomogeneous, and that migration is almost instantaneous, in contradiction with usual observations. This casts doubts on the previous studies on the shear-thinning behavior of these materials and on the concentration dependence of viscosity, and on the interpretations drawn from these macroscopic experiments. However we showed that pointing out problems is not the only use of MRI measurements: combining rheometric measurements and velocity and concentration profiles measurements have allowed us to find the shear part of the constitutive law of the material; we showed that the flowing material behavior is a purely viscous behavior. It also allowed us to perform for the first time local measurements of the concentration dependence of viscosity, based on the real local shear rate and the real local concentration. Our experiments may thus provide a fair experimental basis for flow predictions and for comparison with theoretical models; we find a Krieger-Dougherty dependence $\eta(\phi) = \eta_0(1 - \phi/\phi_m)^{-n}$, of maximum packing fraction $\phi_m = 0.605$ and divergence exponent $n = 2$.

We have proposed a simple constitutive law. The flowing material constitutive law is that of a purely viscous material, and the viscosity follows a Krieger-Dougherty law of maximum packing fraction 0.605 and divergence exponent 2; no flow is allowed for a packing fraction higher than 0.605. The material also exhibits normal stresses proportional to the shear rate. Moreover, there are no steady flows below a critical shear rate $\dot{\gamma}_c$. This behavior accounts for all the macroscopic and local observations: yield stress, shear-thinning and macro viscous behavior, proportionality between shear stress and normal stresses, shear localization at low velocity and high packing fractions, and viscosity bifurcation. In the jammed zone, there must now be a contact network, whereas in the sheared zone there may be only hydrodynamic interactions: localization consists in a change in configuration at the grain scale, which necessitates a deformation of order unity.

These results raise many questions. First, the physical origin of the critical shear rate at the origin of shear localization remains to be investigated. The role of sedimentation should particularly be examined; it seems negligible in our suspensions but gravity (or buoyancy) may nevertheless drive the emergence of the contact network since the particles are very close to each other. It would be interesting to study the influence of a change in the particle-to-fluid density ratio on the localization dynamics. It would also certainly be important to determine the critical shear rate dependence on concentration, which can be done only with accurate MRI measurements (at this stage, we did not have accurate measurements of the shear rate at the interface between the sheared and the unsheared material). We would also need to study the suspensions behavior at very low velocities, for which we expect the flow to be strongly localized as for dry granular materials, in order to test the existence of frictional flows and the condition for their emergence.

We also showed that there still lacks a good modelling of migration in dense suspensions. In order to provide experimental results, we now intend to study the onset of migration. In the diffusive model framework, there is a particular need for experimental results on single particle diffusion so as to confirm the strong dependence of the dimensionless diffusion constant on concentration we inferred from our experiments. Finally, we plan to study experimentally the interplay between localization and migration at low velocities; these experiments will provide a rich test of our understanding of the material behavior.

*G.O. thanks Daniel Bonn, Xavier Chateau, and Joe Goddard for fruitful discussions, and is particularly grateful to Philippe Coussot for his very constructive remarks about the present paper.*